\documentclass[12pt]{iopart}

\usepackage{graphicx}
\usepackage{comment}
\usepackage[colorinlistoftodos]{todonotes}
\usepackage[shortcuts]{extdash}
\usepackage{setspace}
  \expandafter\let\csname equation*\endcsname\relax
  \expandafter\let\csname endequation*\endcsname\relax
\usepackage{amsmath}

\begin{document}
\title{Alloying, de-alloying and reentrant alloying in (sub-)monolayer growth of Ag on Pt(111) }

\author{Maciej Jankowski$^{1,2,\S,*}$, Esther van Vroonhoven$^{1,\S}$, Herbert Wormeester$^1$, Harold~J.~W. Zandvliet$^1$ and Bene Poelsema$^1$}

\address{$^1$ Physics of Interfaces and Nanomaterials, MESA+ Institute for Nanotechnology, University of Twente, P.O. Box 217, 7500AE Enschede, the Netherlands}
\address{$^2$ Present address: ESRF- The European Synchrotron, 71 Avenue des Martys, Grenoble 38043, France}

\address{$^{\S}$ Both authors contributed equally to the presented work.}

\address{$^{*}$ Corresponding author: maciej.jankowski@esrf.fr}

\maketitle 

\begin{abstract}
An in-situ nanoscopic investigation of the prototypical surface alloying system Ag/Pt(111) is reported. The morphology and the structure of the ultrathin  Ag-Pt film is studied using Low Energy Electron Microscopy during growth at about 800 K. An amazingly rich dynamic behaviour is uncovered in which stress relieve plays a governing role. Initial growth leads to surface alloying with prolonged and retarded nucleation of ad-islands. Beyond 50\% coverage de-alloying proceeds, joined by partial segregation of Pt towards the centre of large islands in violent processes. Upon coalescence the irregularly shaped vacancy clusters are filled by segregating Pt, which then take a compact shape (black spots). As a result at around 85\% coverage the strain of the initially pseudo-morphological film is almost completely relieved and Pt-segregation is at its maximum. Further deposition of Ag leads to transient re-entrant alloying and recovery of the pseudo-morphological layer. The black spots persist even in/on several layers thick films. Ex-situ atomic force microscopy data confirm that these are constituted by probably amorphous Pt(-rich) structures.  The (sub-)monolayer films are very much heterogeneous. 

\end{abstract}
\pacs{68.35.Fx, 68.49.Jk, 81.10.Pq}
\maketitle

\section{Introduction}
 
The growth of ultra-thin Ag films on Pt(111) has received much attention in the past few decades~\cite{Davies1982, Paffett1985, Roder1993a, Hartel1993, Grossmann1996, Schuster1996, Becker1993}, caused mainly by complex surface alloying~\cite{Roder1993, Struber1993, Bendounan2012} resulting in the formation of stress stabilized surface nanostructures~\cite{Zeppenfeld1994, Zeppenfeld1995, Tersoff1995, Jankowski2014}. The interest was further raised by the possibility to generate and tune novel chemical and physical properties by varying stoichiometry at the surface and careful control of the Ag growth conditions~\cite{Jankowski2014, Diemant2015, Schuettler2015}, and the formation of periodic dislocation networks~\cite{Brune1994, Ait-Mansour2012, Jankowski2014a, Hlawacek2015} used as nano-templates~\cite{Brune1998} for the growth of organic films~\cite{Ait-Mansour2006, Ait-Mansour2008, Ait-Mansour2009, Ait-Mansour2009a}.   

Low-energy electron diffraction (LEED)~\cite{Paffett1985}, thermal energy atom scattering (TEAS)~\cite{Becker1993} and scanning tunnelling microscopy (STM)~\cite{Roder1993a} experiments revealed that at room temperature the first Ag layers grew through the formation of large pseudomorhic and thus strained Ag islands. This strain is caused by an about 4\% lattice mismatch between the lattice constants of bulk Ag and Pt. An increase of the surface temperature above 550~K leads to irreversible disorder at the surface~\cite{Becker1993}. STM investigations~\cite{Roder1993} revealed that this disorder is caused by the formation of a surface confined alloy~\cite{Tersoff1995} comprised of strained nanometre-sized Ag-rich structures embedded in the Pt surface~\cite{Roder1993}. The shape and size of these structures varied strongly with coverage~\cite{Zeppenfeld1994, Zeppenfeld1995, Jankowski2014}. After deposition of one monolayer~(ML) of Ag the surface was found to dealloy and a pseudomorphic Ag layer was found. Further deposition of Ag induced the formation of a triangular dislocation network which allows relief of the surface strain~\cite{Brune1994}. The third layer was reported to be a pure silver layer~\cite{Ait-Mansour2012} with a propagation of a height undulation originating from the dislocation network at the buried interface~\cite{Ait-Mansour2012, Jankowski2014a, Hlawacek2015}.
 
Much important microscopic work on the growth of ultra-thin silver films on Pt(111) has been done since this system has become a prototype of surface confined alloying.  However, in-situ spatio-temporal information during deposition is still completely lacking. The present study fills this remarkable gap. As will be shown the system shows incredibly rich dynamic behaviour which provides important details of the (de\=/)alloying processes in this system. Our investigation reveals that during growth of the first layer at 750-800~K a AgPt surface alloy forms with areas exhibiting a different AgPt-stochiometry. Beyond about 0.5~ML added Ag, the surface starts to dealloy which becomes prominent near coalescence of the first layer islands. Further increase of the Ag coverage results in the formation of irregularly shaped vacancy islands which are gradually filled by Pt atoms expelled from the alloy phase. In the last stage, these segregated Pt atoms form compact clusters which are observed as black spots in low energy electron microscopy (LEEM) images. These areas have also been analysed \textit{ex~situ} with atomic force microscopy (AFM). The combination of these techniques leads to our identification of these black spots as heavily strained and possibly even amorphous Pt features. When the coverage reaches 1~ML we observe that the integral area of these clusters decreases, which is attributed to their partial dissolution caused by a reentrant alloying of the ultrathin Ag-film. These ``amorphous'' Pt clusters are visible on the surface up to a coverage of several layers and are stable upon exposure to atmospheric conditions.

\section{Experimental}

The LEEM measurements were performed with an Elmitec LEEM~III with a base pressure better than $1 \times 10^{-10}$~mbar. The used Pt(111) crystal had a miscut angle of less than 0.1$^{\circ}$~\cite{Linke1985}. Surface cleaning was done by prolonged repetitive cycles of argon ion bombardment, annealing in oxygen at $2 \times 10^{-7}$~mbar at 800~K, and subsequent flashing to 1300~K in the absence of oxygen. The sample was heated by electron bombardment from the rear and the temperature was measured with a W3\%/Re-W25\%/Re thermocouple. An Omicron EFM-3 evaporator was used to deposit 99.995\% purity silver from a molybdenum crucible onto the sample. The coverage was calibrated on the basis of recorded LEEM movies, where we could track growth of the first pseudomorphic silver layer followed by the growth of second layer which propagates from step edges. One monolayer (ML) is defined as a layer with an atomic  density equal to that of a Pt(111) layer. LEEM images were recorded in bright-field mode using an appropriate contrast aperture around the (00) diffracted beam in the focal plane. The low energy electron diffraction (LEED) patterns were recorded using the largest available aperture of 25~$\rm\mu$m in the illumination column. 

The AFM measurements were done under ambient conditions with an Agilent~5100 AFM employing amplitude-modulation for recording height topography. A MikroMasch Al-black-coated NSC35 Si$_{3}$N$_{4}$ AFM tip with a tip radius of 8~nm was used in these measurements. The resonance frequency of this tip was 205~kHz and the nominal spring constant 8.9~N/m. For the measurements an amplitude set-point of 90\% was used and an oscillation amplitude in the range from 30 to 40~nm. The amplitude modulation imaging mode provides  simultaneously topographic and phase images. The latter provides information on the local variation in energy dissipation involved in the contact between the tip and the sample. Various factors are known to influence this energy dissipation, among which are viscoelasticity, adhesion and chemical composition~\cite{Garcia2007}.

\section{Results and Discussion}
\subsection{An overview of initial growth and alloying of Ag on Pt(111); LEEM}

\begin{figure*}[h!] 
\centering
\includegraphics[scale=0.6]{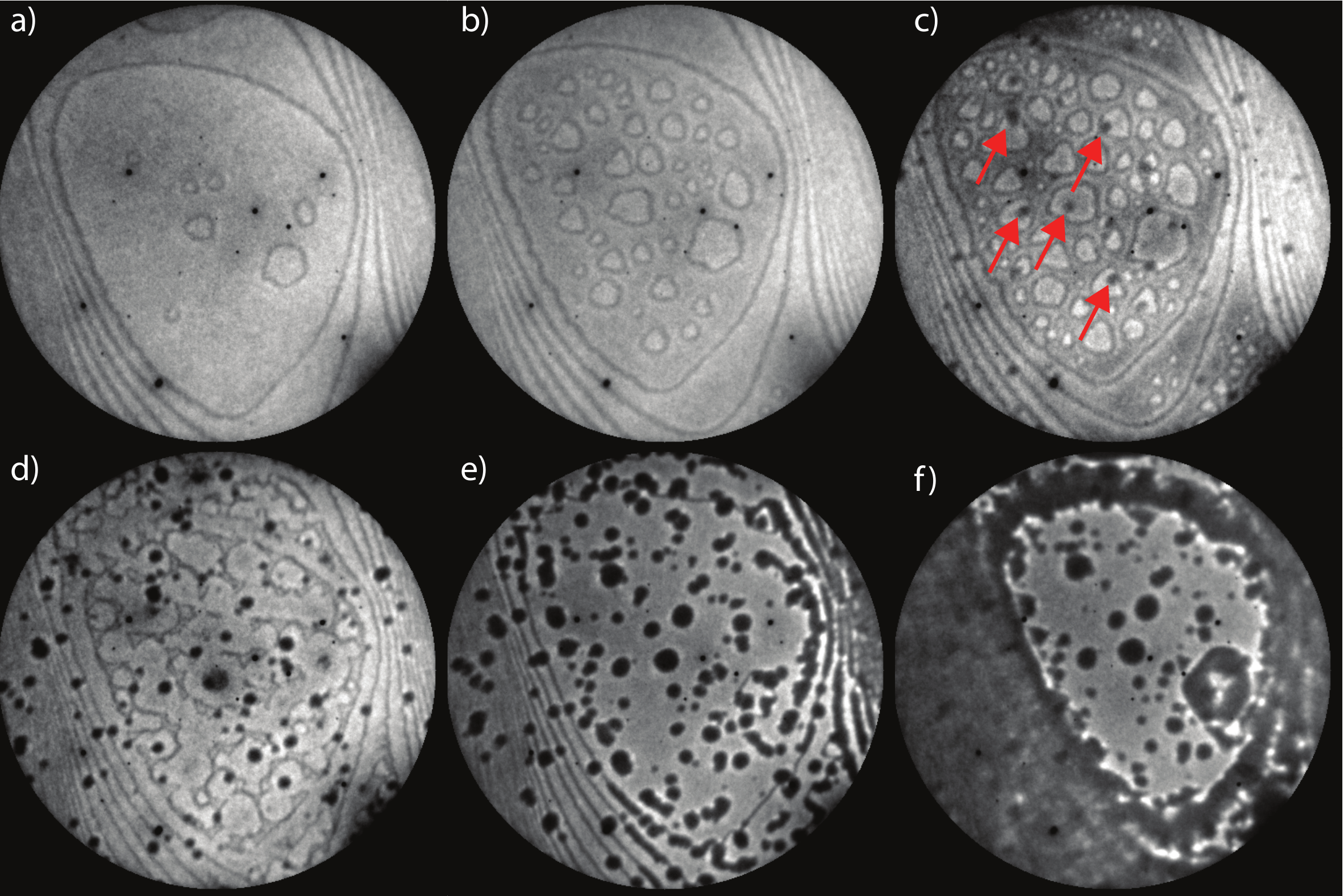}
\caption{LEEM images recorded in bright-field mode at coverage:(a)~0.275~ML, (b)~0.38~ML, (c)~0.5~ML, (d)~0.76~ML, (e)~1 ML, (f)~1.25~ML. The red arrows in (c) points emerging ``black'' spots. The FoV is 2~$\rm{\mu}$m, the electron energy is 2~eV and the substrate temperature is 750~K. The small dark spots, best visible in fig.~(a) are LEEM channel plate defects.}
\label{fig:leem_frames_alloying}
\end{figure*}

Figure~\ref{fig:leem_frames_alloying} shows a sequence of LEEM images recorded during the growth of the first Ag layer on Pt(111) at 750~K. The brightness of the various features in these successively recorded images cannot be compared due to the digital enhancement of each individual image in order to obtain always the best contrast settings. An image with a 2~$\rm\mu$m field of view (FoV) of the initial Ag deposition on Pt(111) surface is shown in Fig.~\ref{fig:leem_frames_alloying}(a). The results have been obtained with a deposition rate of $1.7\times10^{-3}$~ML/s. Figure~\ref{fig:leem_frames_alloying}(a) shows a large terrace in the centre, bordered by ascending monoatomic steps. Upon deposition the growth of a surface confined alloy is observed, starting from the ascending step edges~\cite{Roder1993a} and followed by the nucleation and growth of alloyed islands seen in Fig.~\ref{fig:leem_frames_alloying}(a)\=/(b). Beyond a Ag coverage of roughly 0.5~ML [Fig.~\ref{fig:leem_frames_alloying}(c)] the surface dealloys~\cite{Zeppenfeld1994} slowly and the islands start to coalesce. During the dealloying phase the surface is quite heterogeneous, especially but not exclusively, on top of the largest islands. Darkish or ``black'' spots [marked by red arrows in Fig.~\ref{fig:leem_frames_alloying}(c)] develop first in the centre of the adatom islands and later at an enhanced rate during coalescence. The black dots reveal Pt segregation upon which we focus on in this paper, not only on the process but also on its physical background ascription. At 0.75~ML, coalescence of the islands leads to formation of elongated vacancy clusters [meandering ``lines'' in Fig.~\ref{fig:leem_frames_alloying}(d)]. With continuing Ag deposition these vacancy clusters are filled by expelled Pt atoms and at 1 ML [Fig.~\ref{fig:leem_frames_alloying}(e)] they then take compact shapes as we will detail further below. Completion of the monolayer leads to a decrease of the integral exposed area of the black dots. This shrinkage of the dots is accompanied with reduced total brightness and indicates re-entrant alloying. The second layer starts to grow by step-propagation [see Fig.~\ref{fig:leem_frames_alloying}(f)] and island nucleation in the center of terraces. 

\begin{figure*}[h!] 
\centering
\includegraphics[scale=0.75]{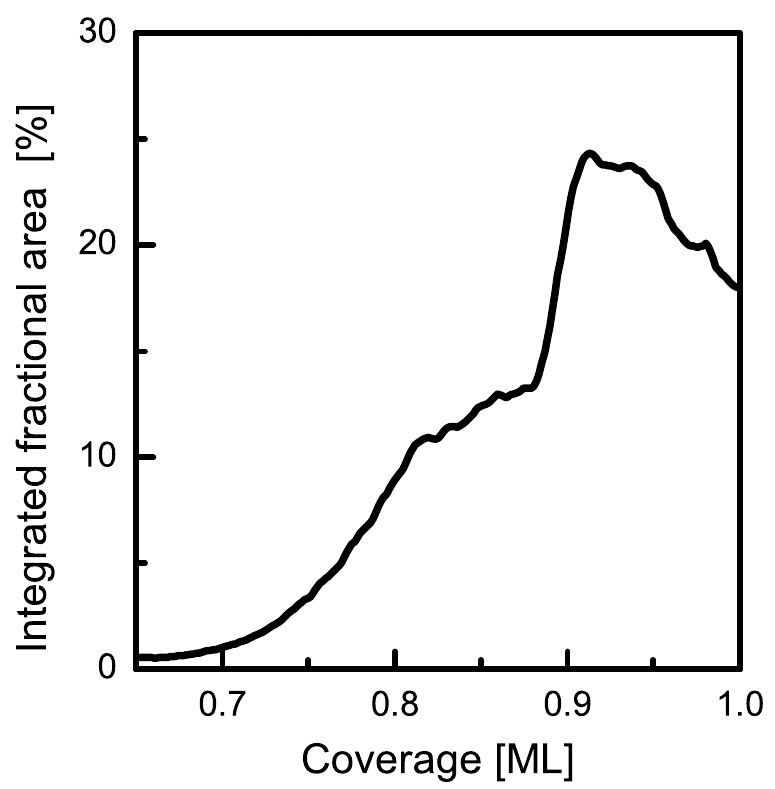}
\caption{Integrated fractional area of the black spots~vs~coverage. The fractional area is calculated from the images recorded with 4~$\rm{\mu}$m FoV.}
\label{fig:black_spots_area}
\end{figure*}

To illustrate the evolution of the Pt segregation we plot in Fig.~\ref{fig:black_spots_area} the fractional area of the black spots as a function of the coverage. The initial increase of the area, up to 0.8~ML, is caused by development of the black spots in the centres of the islands. Later, the steep increase, at around 0.85~ML, is related to the filling of the vacancies by Pt atoms at an enhanced rate, which is followed by a distinct decrease just before completion of the first layer. This decrease reveals the partial dissolution of the black spots due to the re-entrant alloying of the film when approaching completion of the first monolayer.

\subsection{Mixing in initial sub-monolayer growth of Ag on Pt(111) ; LEEM}

\begin{figure*}[h!] 
\centering
\includegraphics[scale=0.75]{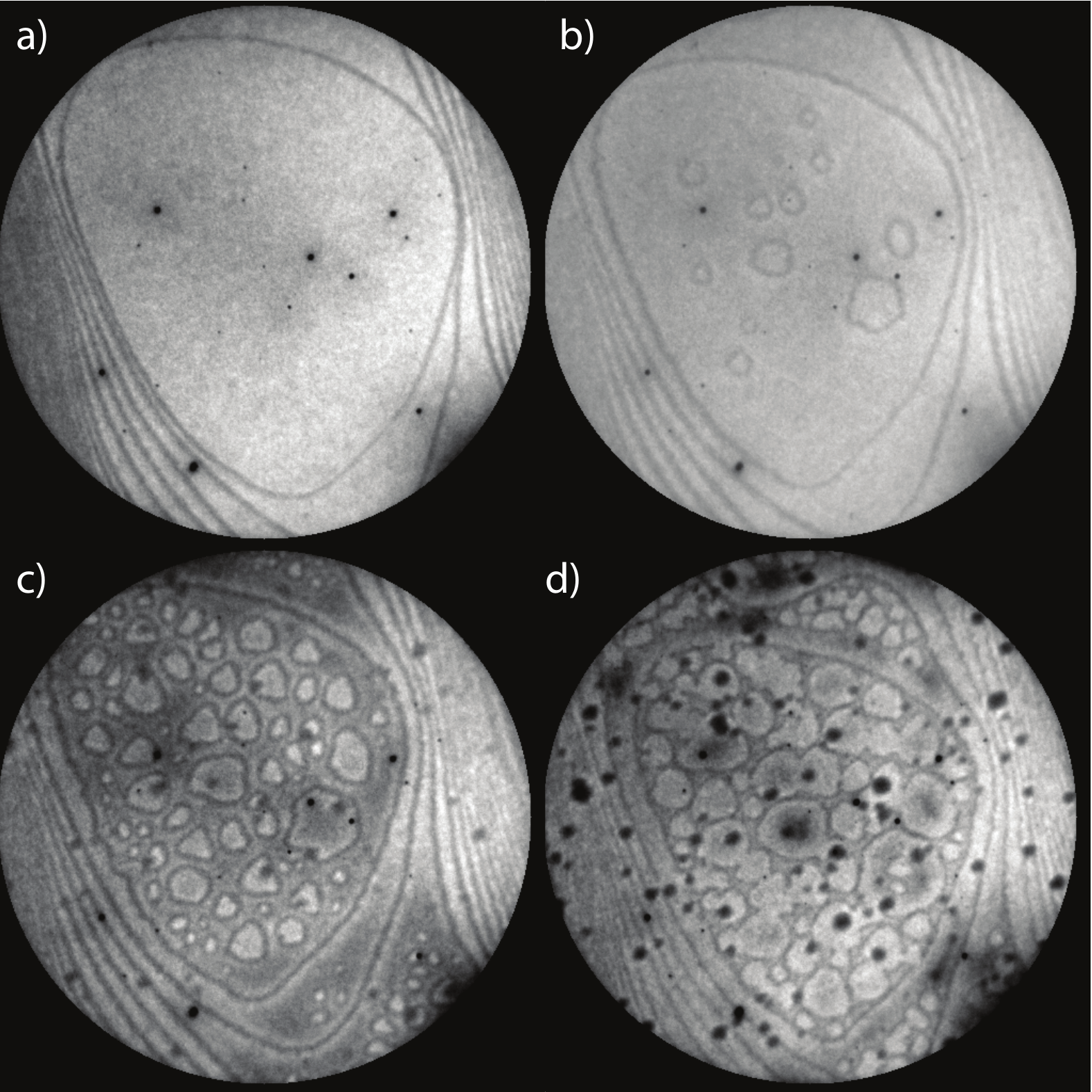}
\caption{LEEM images recorded in bright-field mode :(a)~clean~Pt(111), (b)~0.3~ML, (c)~0.48~ML, (d)~0.7~ML. The FOV is 2~$\rm{\mu}$m, the electron energy is 2.0~eV and the substrate temperature is 750~K. The small dark spots, best visible in fig.~(a) are LEEM channel plate defects.}
\label{fig:esther_initial}
\end{figure*}

Figure~\ref{fig:esther_initial} shows representative snapshots of a movie taken during the initial growth of Ag/Pt(111). The brightness of the exposed layer decreases immediately after opening of the Ag evaporator shutter. This is indicative of an increase of the diffuse scattering due to alloying related disorder~\cite{Becker1993} and will be discussed in more detail in the next section. It is also obvious that at first the material is accommodated at the ascending steps. These steps propagate downward toward the large terrace in the center. Initially effectively all material, i.e., the excess Ag atoms and the Pt atoms expelled from the exposed surface are sufficiently mobile to reach the pre-existing steps. During these initial stages the brightness of the surface is identical everywhere indicating that the emerging surface undergoes alloying in which the Ag embedded in the surrounding Pt matrix~\cite{Roder1993,Zeppenfeld1994,Zeppenfeld1995} is homogeneously distributed. There is also no brightness variation on the narrower terraces near the outer skirts of the images. After some incubation time of about 180~s (0.3~ML) a few islands, visible in Fig.~\ref{fig:esther_initial}(b), nucleate in the central region of the large lower terrace. This provides another indication of a peculiar, alloying related feature. Where initially the mobility was sufficiently large for the atoms to reach the bordering steps, the actual mobility becomes insufficient to reach the propagating steps even if their separation has decreased. This is the first indication for a decreasing mobility of (Pt- and Ag) ad-atoms on the alloying surface. Further evidence for this process is obtained upon a comparison of Figs.~\ref{fig:esther_initial}(c)-(d). 

\begin{figure*}[h!]
\centering
\includegraphics[scale=1.3]{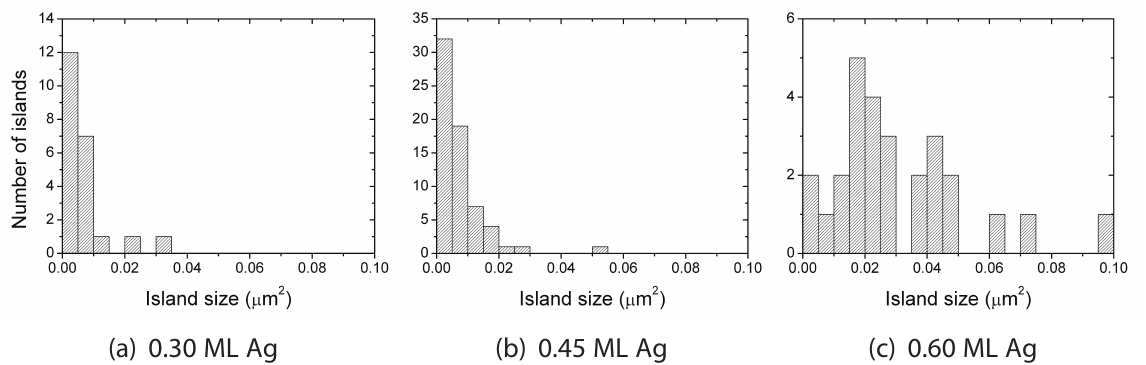}
\caption{The island size distribution for various submonolayer coverages of 0.30, 0.45 and 0.60 monolayers presented in (a), (b) and (c), respectively}
\label{fig:esther_graph_islands}
\end{figure*}

The number of islands increases up to very late stages of growth as illustrated in Fig.~\ref{fig:esther_graph_islands}. Unlike in conventional nucleation and growth where nucleation is about finished after deposition of 1\% of a monolayer and the emerging islands keep growing with a quite narrow size distribution, until they coalesce~\cite{Venables2000}. In the present case nucleation still is active at~≈~0.60~ML. Moreover, the island size distribution is extremely broad and positional distribution is all but homogeneous. The smallest islands emerge near the atomic steps in the originally denuded zone. 

\begin{figure*}[h!] 
\centering
\includegraphics[scale=0.75]{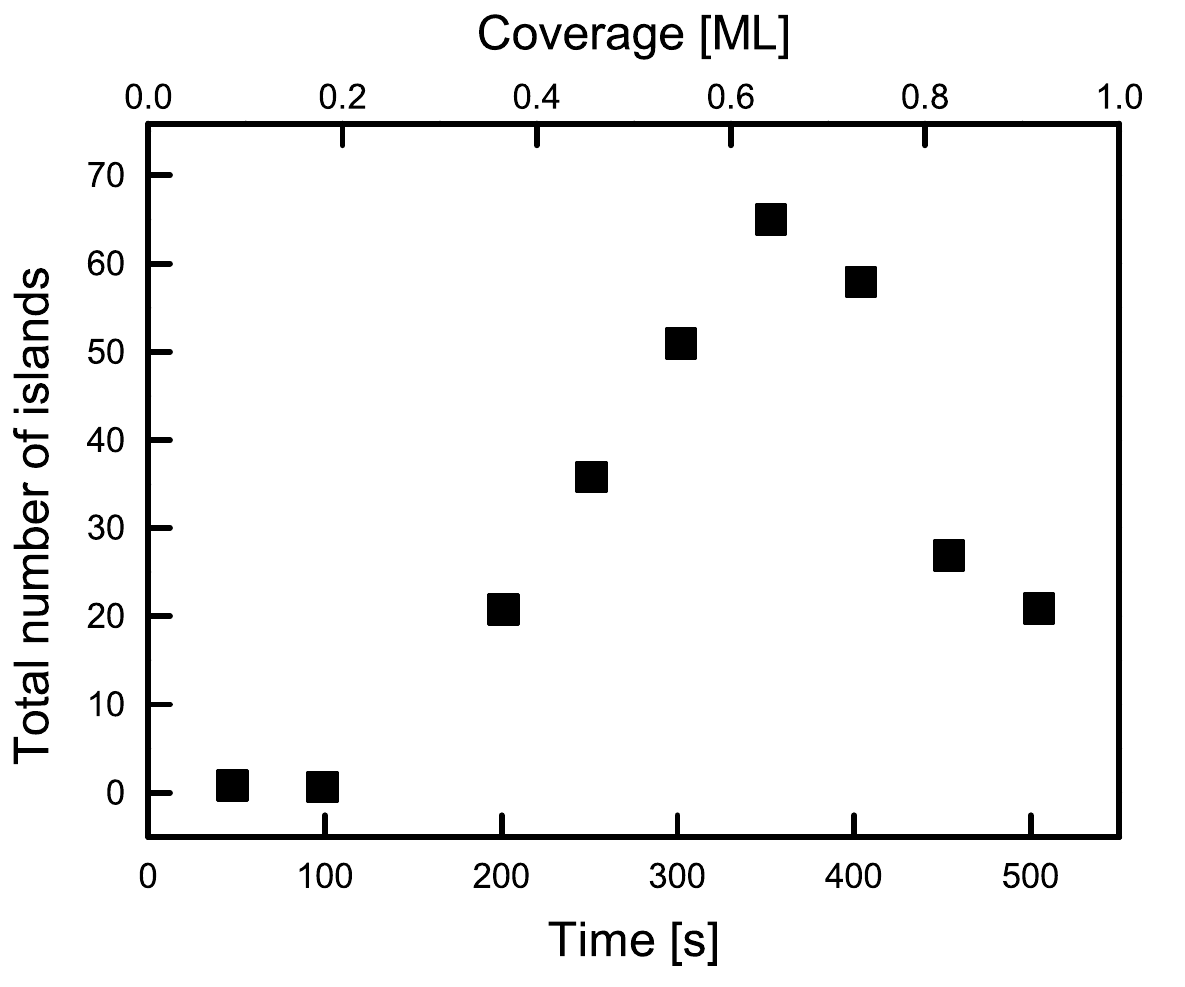}
\caption{Number of first layer islands as a function of deposition time. The deposition rate is $1.7\times10^{-3}$~ML/s.}
\label{fig:esther_graph_N_islands}
\end{figure*}

Another way of illustrating the late nucleation behaviour is depicted in Fig.~\ref{fig:esther_graph_N_islands}, where we plotted the number of islands as a function of time. Progressive nucleation takes place, which leads to a peak in the island density at 64\% of a monolayer. A straightforward explanation for these combined observations is alloying in the two exposed levels. Due to the progressive heterogeneity of the top layer the effective diffusion length decreases continuously and dramatically. As a result, enhanced nucleation takes place. In a site hopping framework one would conclude that the activation energy for diffusion increases by not less than a factor of 4. One safely arrives at the conclusion that initially the mobility of the ad-atoms decreases with increasing coverage. However, an attempt to nail this down to more definite numbers will fail due to the complexity of the system. The diffusing species are both Ag atoms and expelled Pt atoms and these move across a heterogeneous top-layer with small Ag-rich patches embedded in the Pt matrix. Moreover, with increasing coverage the concentration of embedded Ag atoms increases, while in a site hopping model the residence times on top of Ag filled sites and on top of Pt filled sites will most likely be different for diffusing Ag atoms and Pt atoms. One cannot exclude either that exchange processes further slow down the diffusion rates. We note that the de-mixing occurring at higher coverage leads to a more homogeneous composition and, as a consequence of reversing the argument, to enhanced diffusion and thus exclusion of new nucleation events. During these late stages of monolayer growth the number of islands also decreases as a result of coalescence processes.

\subsection{De-alloying of the alloy above 0.5 ML ; LEEM}

\begin{figure*}[h!] 
\centering
\includegraphics[scale=0.75]{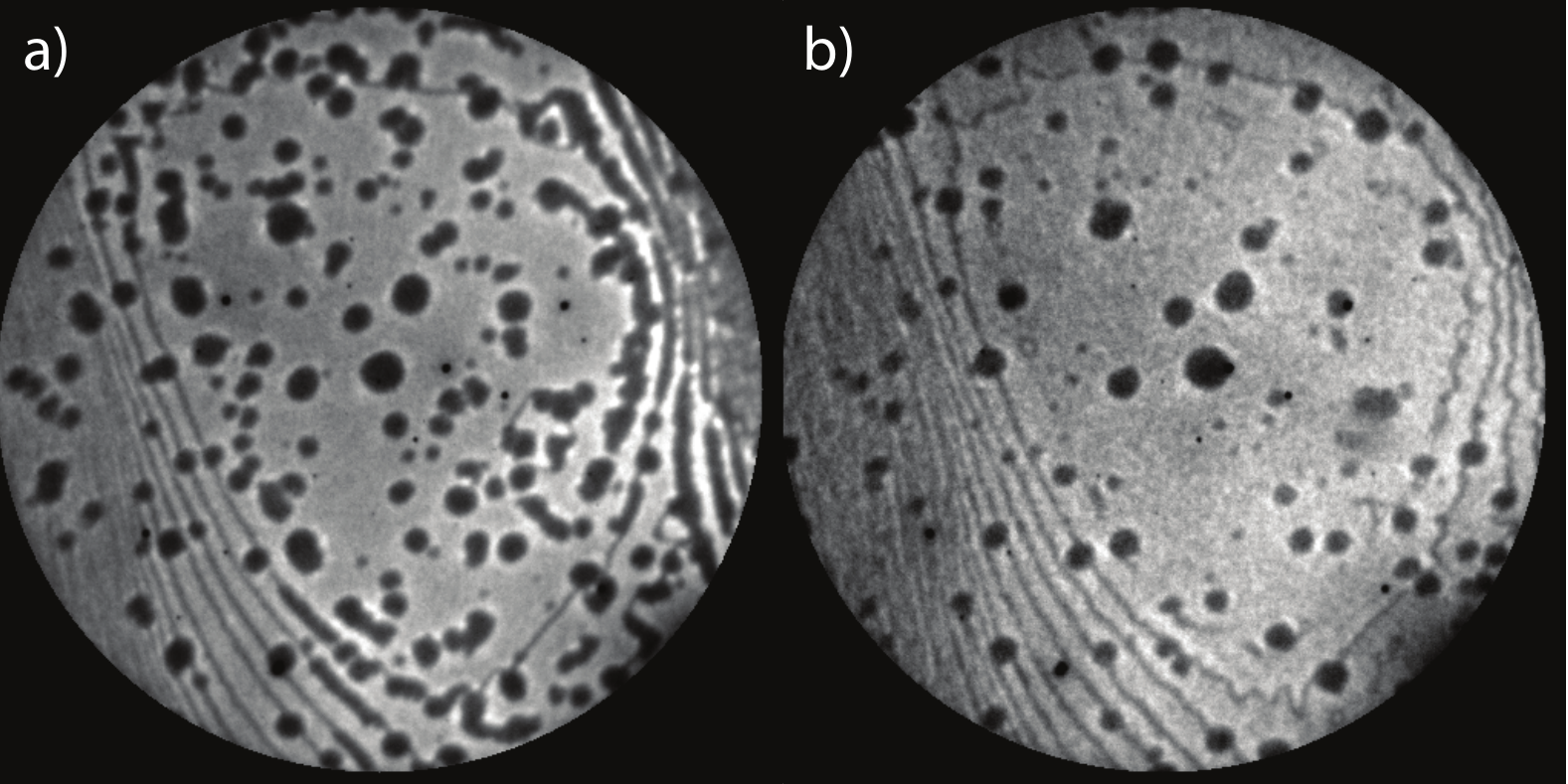}
\caption{LEEM images recorded in bright-field mode :(a)~1.0~ML, (b)~1.68~ML. The FOV is 2~$\rm{\mu}$m, the electron energy is 2.0~eV and the substrate temperature is 750~K.}
\label{fig:esther_spots_higher_ML}
\end{figure*}

A striking feature is the emergence of black spots as shown in Fig.~\ref{fig:esther_spots_higher_ML}(a)-(b). The black spots do appear when approaching completion of the monolayer and are emerging in Fig.~\ref{fig:esther_initial}(c). They persist after the growth of several monolayers, and are even visible after the completion of the eighth monolayer (not shown here). We have carefully tried to identify their origin. For this purpose we have varied the deposition rate between $1.7\times10^{-4}$ and $1.1\times10^{-2}$~ML/s. Both the number density of black dots and their integrated area are at variance with the anticipated behaviour for contamination: both increase a few tens of a percent, i.e., much less than the factor 13, which might be anticipated on the exposed time of the, compared to Ag, by far more reactive Pt\=/surface layer. Occasionally, we do observe some evidence of carbon impurities (not visible in presented LEEM figures). However, their signature as a result of creating under-focus and over-focus conditions differs completely from that of the black dots in Fig.~\ref{fig:esther_initial}~and~Fig.~\ref{fig:esther_spots_higher_ML}. We therefore conclude that the ``black spots'' are inherent to the Ag-Pt(111) de-mixing process. They appear to consist of disordered Pt-rich patches which reduce in integral size with increasing film thickness, but persist even after deposition of 8 ML.

\begin{figure*}[h!] 
\centering
\includegraphics[scale=0.75]{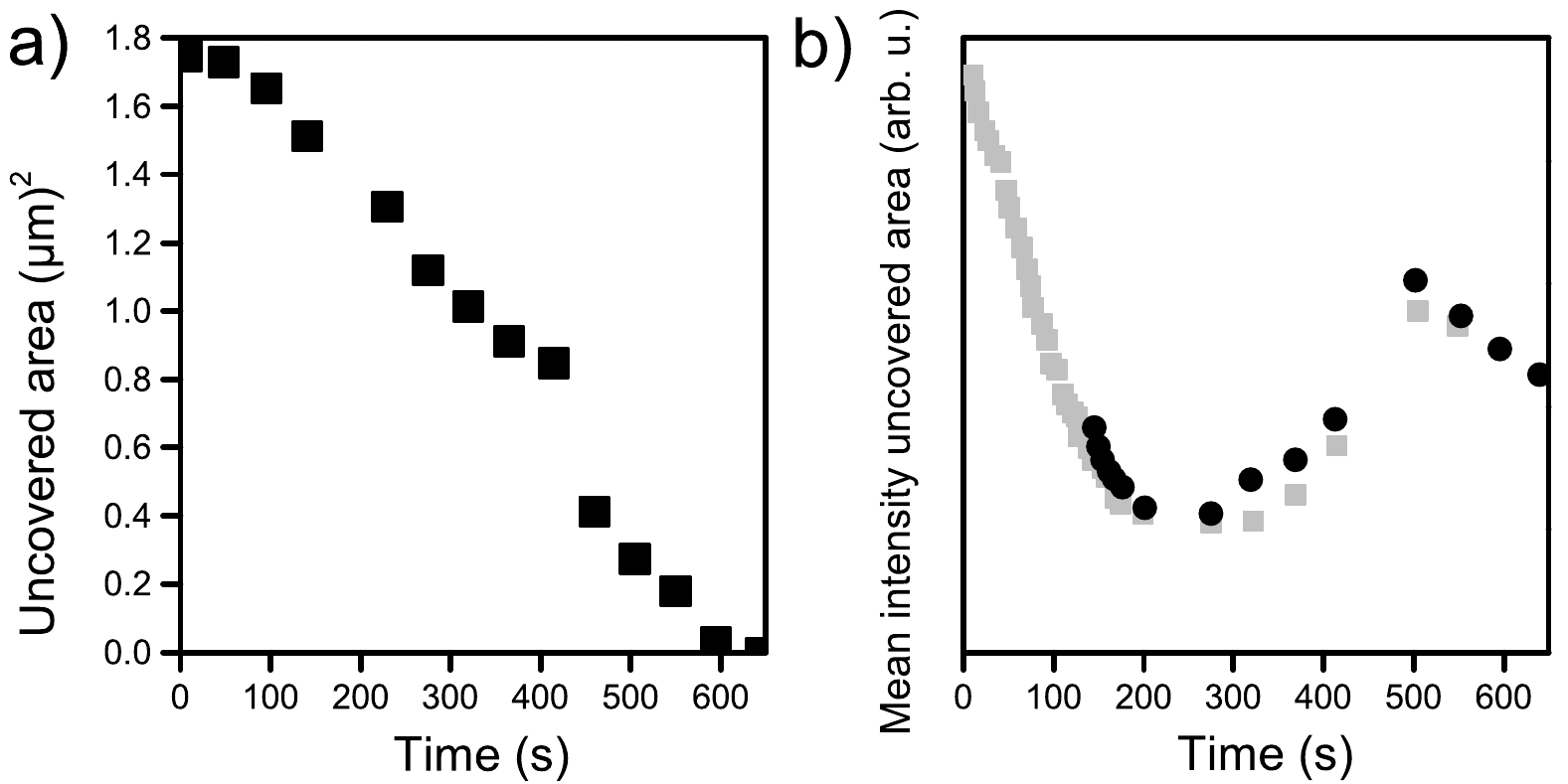}
\caption{(a) the growth process of Fig.~\ref{fig:esther_initial} analysed in the uncovered area (see text). (b) the mean bright field intensity as a function of time. The exposed toplayer of the substrate is represented by the squares, the exposed first layer by circles. The area taken by black spots is subtracted from the analysis. The break above 400 s is caused by not-well defined morphology of the surface caused by rapid de-alloying which causes big uncertainties in the perfomed analysis.  Imaging electron energy is 2 eV.}
\label{fig:esther_area_plot}
\end{figure*}

Figure~\ref{fig:esther_area_plot}(a) shows the evolution of what naively would be called the uncovered area, i.e., the supposedly still exposed virgin Pt(111) surface. As we will see shortly this conjecture cannot be maintained. The area taken by the black dots is subtracted. Within the uncertainties originating from this subtraction and/or from lensing effects as a result of field irregularities due to e.g. work function differences, the ``exposed Pt'' area decreases linearly with time, indicative of the growth of a single layer. In this particular experiment it takes about 590 s to complete the first layer. More information is obtained from the averaged brightness of the layers evolving at the two exposed levels as shown in Fig.~\ref{fig:esther_area_plot}(b). In this evaluation the black dots have been disregarded. The grey symbols refer to the original level of the clean Pt(111) surface, while the black circles represent the results for the exposed surface at a coverage that is one monolayer higher. As becomes immediately evident the brightness decays strongly from the very start of the deposition. It passes through a pronounced minimum and subsequently increases again up to a maximum value after roughly 500~s~(0.85~ML) (as we will explain below this maximum brightness marks busy activity in de-mixing, segregation, stress relaxation and re-entrant mixing). It is also obvious that the average brightness of both exposed layers is very similar. Both observations are completely consistent with the formation of a mixed ad-layer, or the surface confined alloy discussed in Ref.~\cite{Roder1993,Zeppenfeld1995,Zeppenfeld1994}. Alloying proceeds during the deposition of the first half layer, while de-alloying occurs during deposition of the second half monolayer. The alloy can only persist in the vacuum exposed layer(s) due to the involved energetics. First the tensile tension in the Pt(111) is reduced by the incorporation of the larger Ag atoms~\cite{Tersoff1995}, while with increasing Ag content the tension finally becomes compressive, which leads to the exchange of first layer Pt with Ag embedded in the lower layer. In the initial phase the brightness decays due to increasing disorder by the continuing embedding of Ag patches in the Pt\=/matrix, while later on the de-mixing leads to lesser disorder: the continuing expulsion of Ag leads to larger Pt patches surrounding the embedded Ag. 

\begin{figure*}[h!] 
\centering
\includegraphics[scale=1.5]{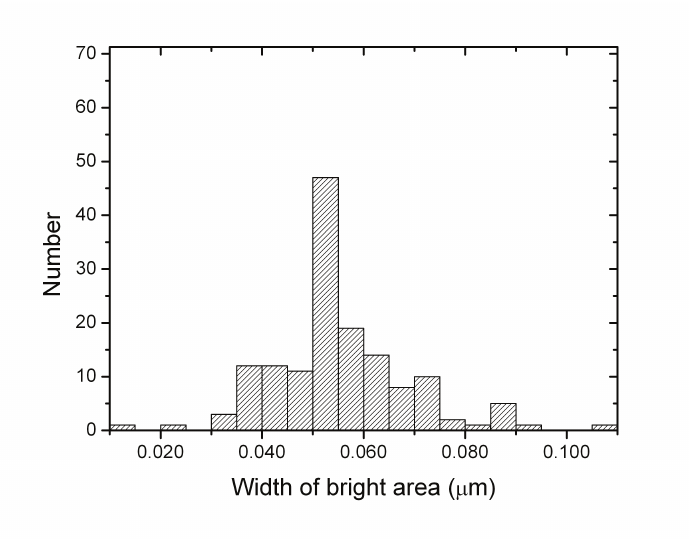}
\caption{The width of the bright border around the large islands has
been measured of a number of islands of several deposition experiments.
There is a strongly proffered.}
\label{fig:esther_rims_plot}
\end{figure*}

The brightness across the level of the growing ad-layer is by no means homogeneously distributed during the growth beyond about 0.40 ML. A representative snapshot is shown in Fig.~\ref{fig:esther_initial}(c). At the chosen imaging conditions the smaller islands are bright, the ones with ``intermediate'' sizes are slightly less bright, many of them have a distinct black dot, and some of them have a distinctly brighter rim, when compared to their centres. This feature is even clearer for the largest islands, which all have a bright rim and a dimmer central part. We interpret these findings as follows: the brighter areas represent Ag(-rich) areas. The smaller islands which appear during late stages of the nucleation are constituted by Ag atoms which are still continuously deposited and Ag atoms which are now expelled from the initial level, i.e., from the alloyed exposed parts of the substrate. During later stages of growth of the intermediately sized and largest islands, the edges also accommodate mostly Ag-atoms for the same reason. The resulting bright rims appear to have a characteristic width. The results obtained for a number of islands from several deposition experiments is shown in Fig.~\ref{fig:esther_rims_plot}. The distribution of the widths of these bright rims/borders is somewhat skewed but has a pronounced maximum at about 52~nm. The fact that these widths show no distinct correlation with island size or deposition rates suggests their thermodynamic origin. We attribute this to the relaxation of stress. The compressed Ag-film can best (partly) relieve its stress near the descending step edges. The emerging of bright rims is responsible for the slightly higher brightness obtained for the highest layer around 300~-~400~s in Fig.~\ref{fig:esther_area_plot}(b). 

\begin{figure*}[h!]
\centering
\includegraphics[scale=0.75]{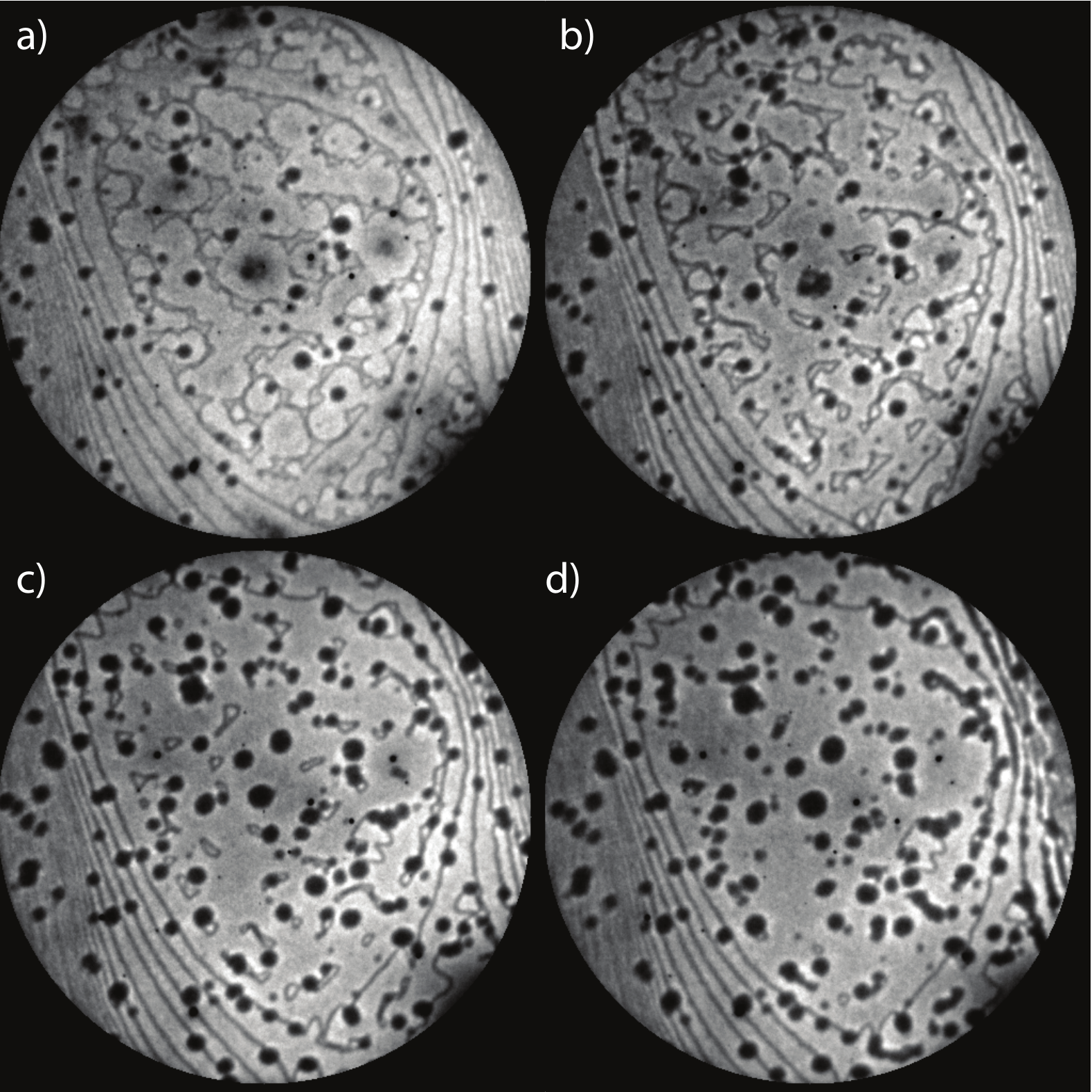}
\caption{LEEM images recorded in bright-field mode :(a)~0.75~ML, (b)~0.82~ML, (c)~0.88~ML, (d)~0.95~ML. The FOV is 2~$\rm{\mu}$m, the electron energy is 2.0~eV and the substrate temperature is 750~K.}
\label{fig:esther_last_stage}
\end{figure*}

\begin{figure*}[h!] 
\centering
\includegraphics[scale=0.75]{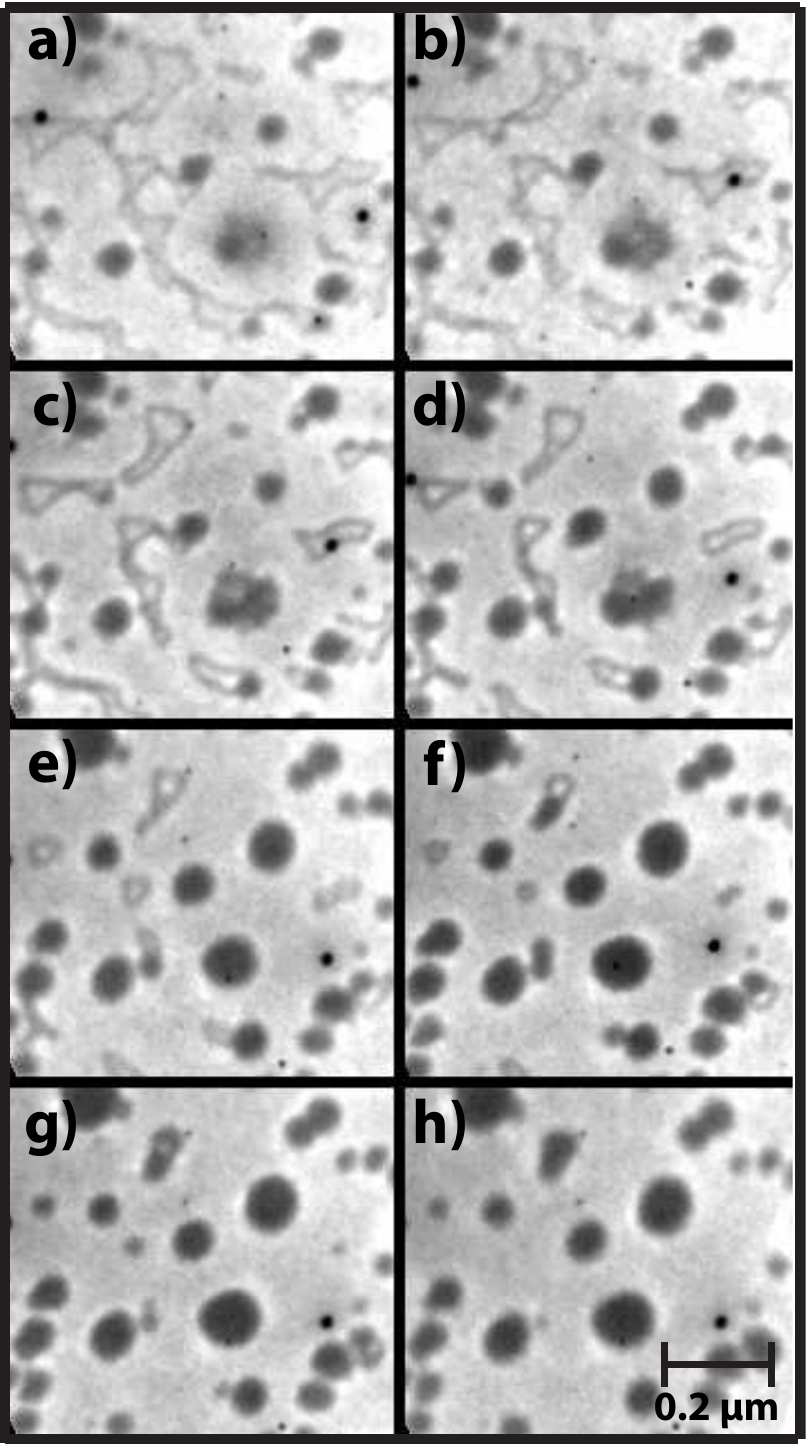}
\caption{LEEM images recorded in bright-field mode at coverage: (a)~0.75~ML, (b)~0.79~ML, (c)~0.82~ML, (d)~0.85~ML, (e)~0.89~ML, (f)~0.92~ML , (g)~0.96~ML,  (h)~1~ML. The electron energy is 1.9~eV }
\label{fig:fig_03_bs}
\end{figure*}

\begin{figure*}[h!] 
\centering
\includegraphics[scale=0.75]{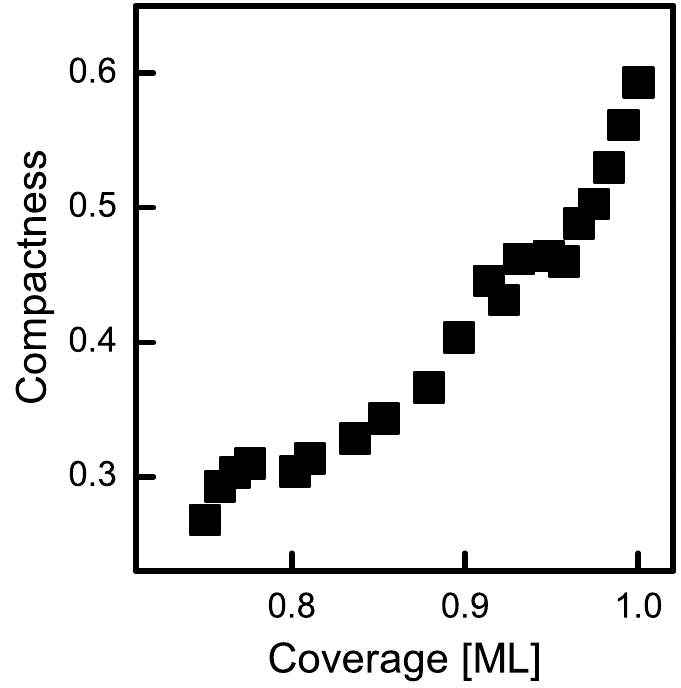}
\caption{Plot of vacancies compactness as a function of coverage.}
\label{fig:fig_04_bs}
\end{figure*}

With increasing silver deposition the de-mixing of the surface confined Ag/Pt alloy continues. The expulsion of surplus Ag from the lower level, i.e., the original substrate level, is apparently facilitated by the advancing steps. However, the Pt atoms, encapsulated on the intermediate and in particular on large islands lack that opportunity and experience difficulties to escape. Therefore, these centres remain relatively dark during a prolonged period. Upon proceeding de-mixing also the larger islands become brighter, the segregation intensifies and dark grey spots evolve, signalling segregation. On the largest islands, the dimmer inner parts are also Pt-rich and strong fluctuations occur as evidenced in Fig.~\ref{fig:esther_last_stage}. The dark grey areas fluctuate strongly both in position and also in brightness. Finally also the largest islands host dark grey spots [cf. Figure~\ref{fig:esther_initial}~(d) taken at a coverage of 0.7 ML]. We note that at the same time the brightness of the small islands is unaffected. Some of the dark grey areas evolve into dark spots, while others seem to dissolve again in the Ag-layer. Around a total Ag coverage of 0.85 ML, i.e., during late stages of coalescence an enormous bustle is taking place. The irregularly shaped vacancy clusters, which are the result of the continuing growth of the coalesced islands, are with overwhelming majority filled by segregation of Pt into black dots [see Fig.~\ref{fig:esther_last_stage}~(c)-(d) and Fig.~\ref{fig:fig_03_bs}]. At the same time the brightness of the large islands has now reached its maximum [see Fig.~\ref{fig:esther_area_plot}~(b)]. This late and strikingly active stage of the birth of new Pt-rich dots happens precisely in those areas where the step advancement ceases. This filling of the irregularly shaped vacancy islands with black Pt-dots goes along with a strong reduction of their integrated length, since all dots are circular in shape. The energetically unfavourable long integrated step length apparently facilitates the Pt segregation after which a much reduced domain border length surrounding the circular black spots, has developed. Both features are quantified and illustrated in Fig.~\ref{fig:fig_03_bs} and Fig.~\ref{fig:fig_04_bs} .

We stress that we did carefully look for indications of structures in the recorded $\mu$LEED patterns during Ag deposition, different from the pseudomorphic (1x1) structure and did not observe any evidence for those. At this same time, as we will show in a moment, we observe relaxations of the surface lattice which reaches maximum at 0.85 ML (see Fig.~\ref{fig:leed_spots}), the coverage where the brightness of the islands reaches maximum in Fig.~\ref{fig:esther_area_plot}~(b). We therefore must conclude that these black dots are probably amorphous Pt(-rich?) areas. Further support for this conclusion is provided by their circular shape: The Pt-rich dots are heavily strained in their Ag-rich environment and undergo strong and isostropic, compressive stress. In a later stage some even coalesce and become larger, eventually resuming their circular shape. We observe that the decrease of the size of the vacancies goes along with a change of their shape to a more compact one. This shape evolution can be seen in Fig.~\ref{fig:fig_04_bs} which shows the compactness of the vacancies as a function of coverage. The compactness $C$ is defined as:

\begin{equation*} 
C = \frac{(A\cdot4\pi)^\frac{1}{2}}{P}
\end{equation*}
where $P$ is the perimeter of vacancies and $A$ is their area. 

At 0.85~ML presented in Fig.~\ref{fig:fig_03_bs}(d), the expelled Pt atoms start to segregate into the vacancies, which finally at 1~ML [see~Fig.~\ref{fig:fig_03_bs}(h)] leads to their complete filing. We note two more things: First with increasing compactness the length of the energetically less favourable edges decreases. Also the step energy will be lower in contact with platinum compared to a vacancy. Both features lead to a lowering of the total energy and thus facilitate the Pt segregation. In other words under heavy compressive stress the exposed Pt atoms actually experience an energetic advantage by segregation into the compacting vacancies. We note that, as illustrated in Fig.~\ref{fig:fig_04_bs}, the compactness never seems to reach its ideal value~of~1. We attribute this to difficulties inherent to setting a well defined discrimination level. Uncertainties related to field distortion effects~\cite{Nepijko2001} play a role and a principal problem is caused by the digital nature of the image. With a finite pixel size one cannot define a line with infinite sharpness. As a result the determined $C$ will always be an underestimate of the real compactness. 

\subsection{LEED spots intensity variation and surface lattice relaxation}

\begin{figure*}[h!] 
\centering
\includegraphics[scale=1]{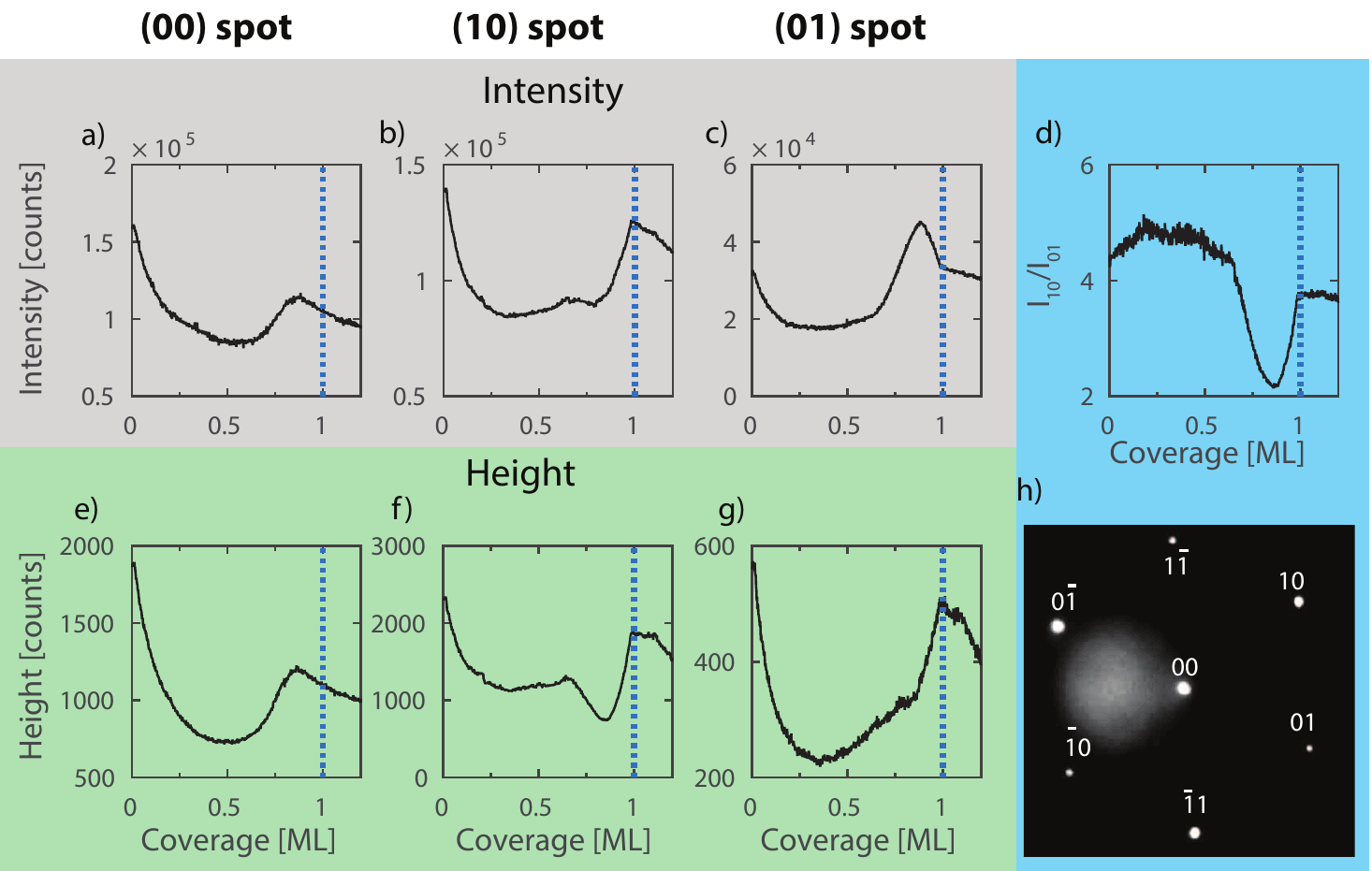}
\caption{Intensity of  LEED spots as a function of coverage: (a)-(00) spot, (b)-(10) spot and (c)-(01) spot. Height of the LEED spots as a function of coverage: (e)-(00) spot, (f)-(10) spot and (g)-(01) spot. Blue line marks completion of the first monolayer. Fig.~(d) represents ratio of the (10) and (01) spots intensity as a function of coverage. Fig.~(h) presents typical LEED pattern recorded from clean Pt(111) with labelled position of corresponding spots. Electron energy is 44~eV, sample temperature is 800~K.}
\label{fig:leed_intensity}
\end{figure*}

\begin{figure*}[h!] 
\centering
\includegraphics[scale=1]{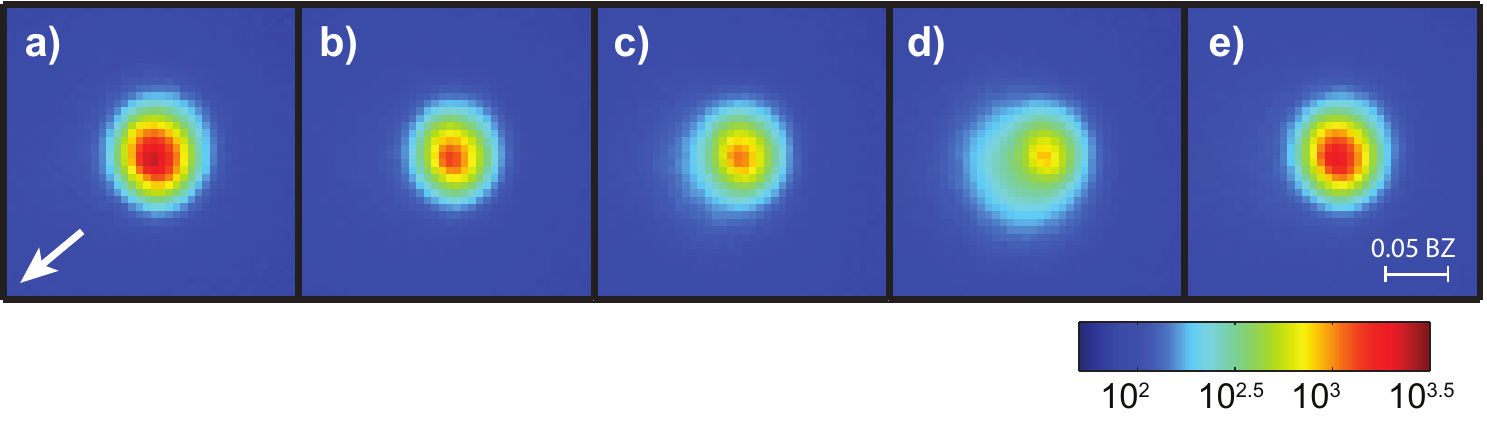}
\caption{The (01) LEED spot recorded during growth of Ag on Pt(111) at 800 K: (a) clean Pt(111), (b) at 0.65 ML, (c) at 0.74 ML, (d) at 0.85 ML, (e) at 1 ML. The arrow in figure (a) points the direction towards the (00) spot. The energy of electrons was 44 eV. The intensity of the spots is presented on logarithmic scale of colours.}
\label{fig:leed_spots}
\end{figure*}

The surface lattice variation can be studied in detail from the evolution of the \textit{in~situ}  $\mu$LEED patterns recorded as function of Ag coverage. In the initial phase of Ag deposition, surface alloying induces the sharp decrease of the intensity of the Bragg spots~\cite{Becker1993,Jankowski2014a} shown in Fig.~\ref{fig:leed_intensity}, but their position corresponding to the Pt(111) lattice [Fig.~\ref{fig:leed_spots}] is not altered. At around 0.7 ML, where coalescence of islands and nucleation of well defined black spots in their centres is observed, the intensity of the (01) spots decreases to reach local minimum at 0.85 ML. At this same coverage, the (01) spot intensity reaches a maximum and both first order spots broaden slightly towards the (00) spot, as shown in Fig.~\ref{fig:leed_spots}(c). We attribute this broadening to the relaxation of the surface lattice towards the 4\% larger Ag distances. This relaxation of the lattice is triggered by de-alloying of the surface and partial segregation of Pt into vacancy clusters as discussed above. Moreover, the value of (10)/(01) spots intensity ratio plotted as a function of coverage in Fig.~\ref{fig:leed_spots}(d) reaches minimum at exactly 0.85 ML. This is a direct result of a transient lowering of the threefold symmetry: The Ag-atoms in the relaxed film not only occupy FCC sites but also HCP sites. The latter contribute to enhanced sixfold symmetry features. 

The first monolayer is pseudomorphic, as its observed p(1$\times$1) LEED pattern given in Fig.~\ref{fig:leed_spots}(e) corresponds to that of the Pt(111) lattice \cite{Paffett1985,Becker1993,Jankowski2014a}. We also mention that after reaching 0.85 ML the intensity of the (00) spot in Fig.~\ref{fig:leed_spots}(a) again decreases, which rationalized as re-entrant mixing of the top layer. This is in line with the partial dissolution of the black spot [cf. Fig.~\ref{fig:black_spots_area}] and the concomitant decrease of the brightness of exposed layers [Fig.~\ref{fig:esther_area_plot}~(b)]              

We now focus on the nature of the black spots and apply different techniques for their characterization. The latter include spatially resolved work function (WF) measurements with LEEM and \textit{ex situ} AFM data. The latter provides support for the conclusion that excreted Pt atoms constitute the black dots. 
\subsection{Work function changes}

\begin{figure*}[h!] 
\centering
\includegraphics[scale=0.9]{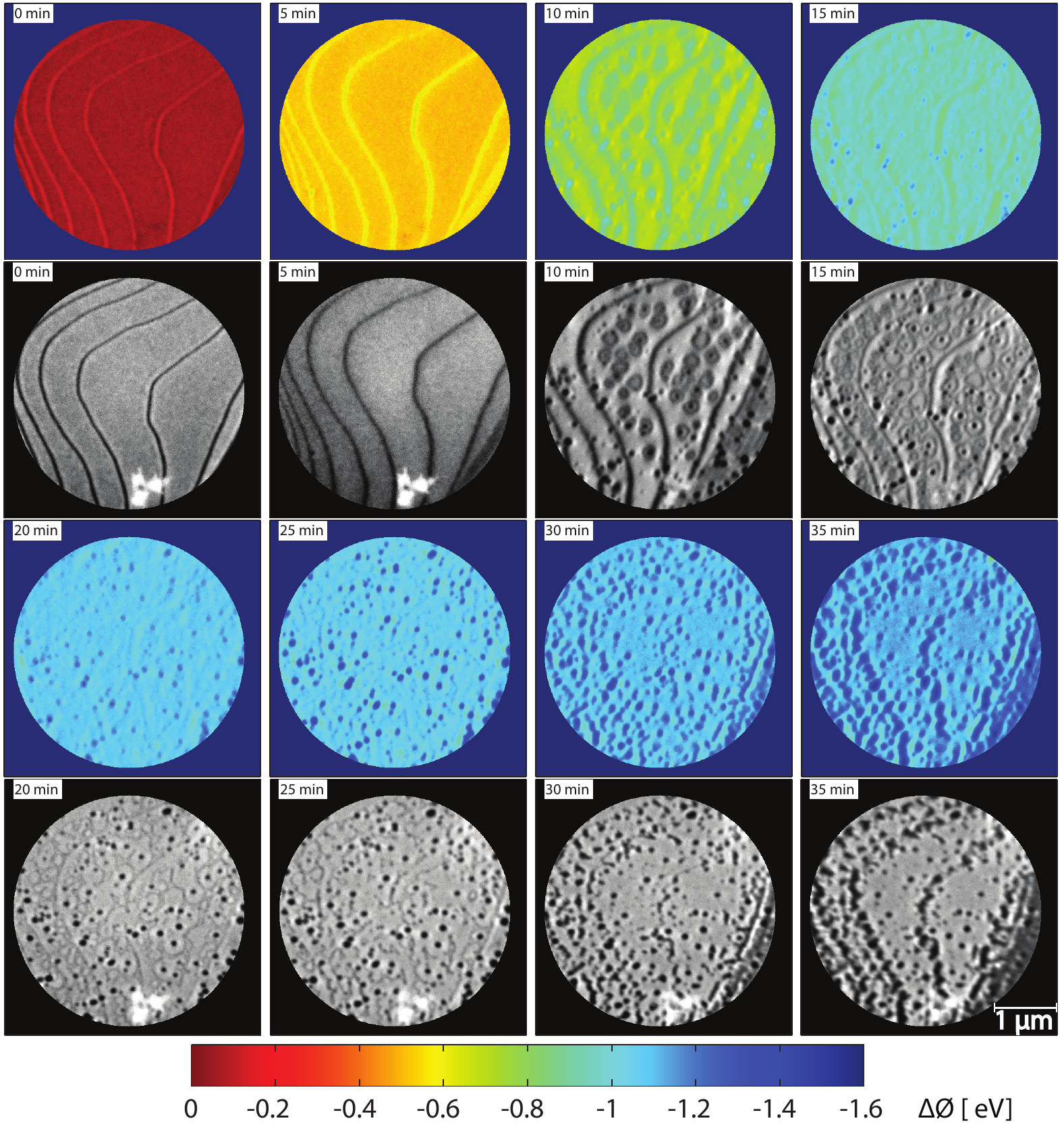}
\caption{LEEM images recorded in bright-field mode at the indicated deposition time of silver (grey scale colour) and the concomitant change in the WF relative to clean Pt(111) (colour scale images). The deposition time needed for completion of the first layer is around $\sim$30~min}
\label{fig:WF_images}
\end{figure*}

\begin{figure*}[h!] 
\centering
\includegraphics[scale=0.75]{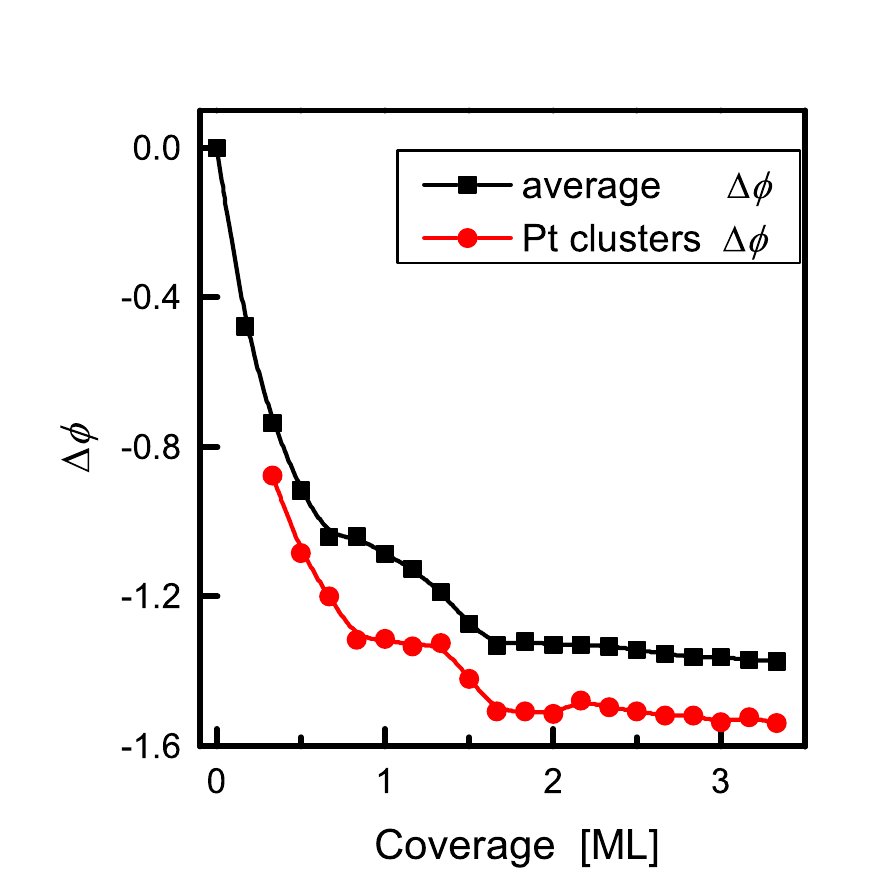}
\caption{Average and Pt clusters work function as a function of coverage.}
\label{fig:fig_07_bs}
\end{figure*}

The LEEM offers a capability of recording spatial maps of relative changes of the surface WF change (see also Ref.~\cite{Hlawacek2015} for more details) during deposition of thin films. Figure~\ref{fig:WF_images} shows a sequence of LEEM images and the corresponding spatial maps of the surface WF change as a function of deposition time.  The atomically clean terraces of Pt(111) exhibit a constant value of the WF and only a 0.2~eV lower WF is seen at the step edges \cite{Besocke1977}. The initial deposition of Ag leads to a sharp decrease of the average WF, shown in Fig.~\ref{fig:fig_07_bs}. After 10~min of Ag-deposition (at around $\sim$0.3~ML) the WF of the growing islands ($\rm{\Delta\phi=-0.79}$~eV) is 0.1~eV lower than WF of the surrounding hosting terrace ($\rm{\Delta\phi=-0.69}$~eV). This contrast in WF is attributed to different levels of intermixing of Ag and Pt in the islands and their surroundings. The work function has been measured also above the black spots, assumed to be Pt(-rich) clusters. As shown in Fig.~\ref{fig:fig_07_bs}, the latter follows quite closely the WF variation of their environment at a distance of about -0.2~eV. The similarity in temporal dependence makes sense if one realizes that the Pt(-rich) clusters are small and long range interactions play a major role in determining the work function. This seems somewhat surprising at first sight since the WF of clean and smooth Pt (111) (5.8~eV) is much larger than that of clean and smooth Ag(111) (4.56~eV)~\cite{Kawano2008}. However, the WF depends strongly on the surface orientation with a tendency for lower WF observation with decreasing coordination, i.e, $\rm{\Delta\phi_{111} < \Delta\phi_{100} <\Delta\phi_{110}}$. This is even worse for amorphous dots with which we most likely deal with. Finite size effects may also lower the WF~\cite{Nepijko2001}. It is noted, that field distortions around small objects make drawing clear conclusions hazardous. Therefore, we tentatively conclude that the case for Pt as the constituting material of the clusters remains undecided on the basis of the current WF data.  

A further increase of the coverage leads to a decrease of the average WF which at coverage of 1 and 2~ML has the value of $\rm{\Delta\phi=-1.085}$~eV and $\rm{\Delta\phi=-1.33}$~eV, respectively. The rather constant value of the WF beyond two layers thick films indicates that at most minor changes in topography and composition of the surface occur beyond a film thickness of two layers.  

At a coverage of 3~ML the average WF saturates at $\rm{\Delta\phi=-1.37}$~eV. This saturation value is in the range reported by Hartel~et~al.~\cite{Hartel1993}~(-1.2~eV) and by Paffet~et~al.~\cite{Paffett1985}~(\=/1.5~eV) obtained at similar coverages. We note that the reported values were measured for the alloy formed after annealing the surface at 600~K. Schuster~et~al.~\cite{Schuster1996} showed, that both temperature and time during alloying drastically influence intermixing levels of the Ag/Pt(111) surface alloy, which results in different surface morphology leading to a variation in the surface WF. Moreover, the step density of the initial Pt(111) surface may have a decisive influence of several tenths of an eV on the measured value \cite{Poelsema1982}. In that respect, we can safely conclude that our data compare favourably with the macroscopic data available in literature. 
 
\subsection{Characterization of 3~ML thick film in ambient AFM}

To gain more insight on the composition and (if possible) structure of the black spots we apply AFM. For that purpose we prepared a 3~layers thick Ag/Pt(111) film by deposition in UHV at 800~K as monitored by LEEM. As evident from Fig.~\ref{fig:WF_images}, and the discussion above, the surface predominantly consists of silver, but a substantial number of black spots is still observed. The AFM data have been taken in ambient atmosphere.

\begin{figure*}[h!] 
\centering
\includegraphics[scale=0.6]{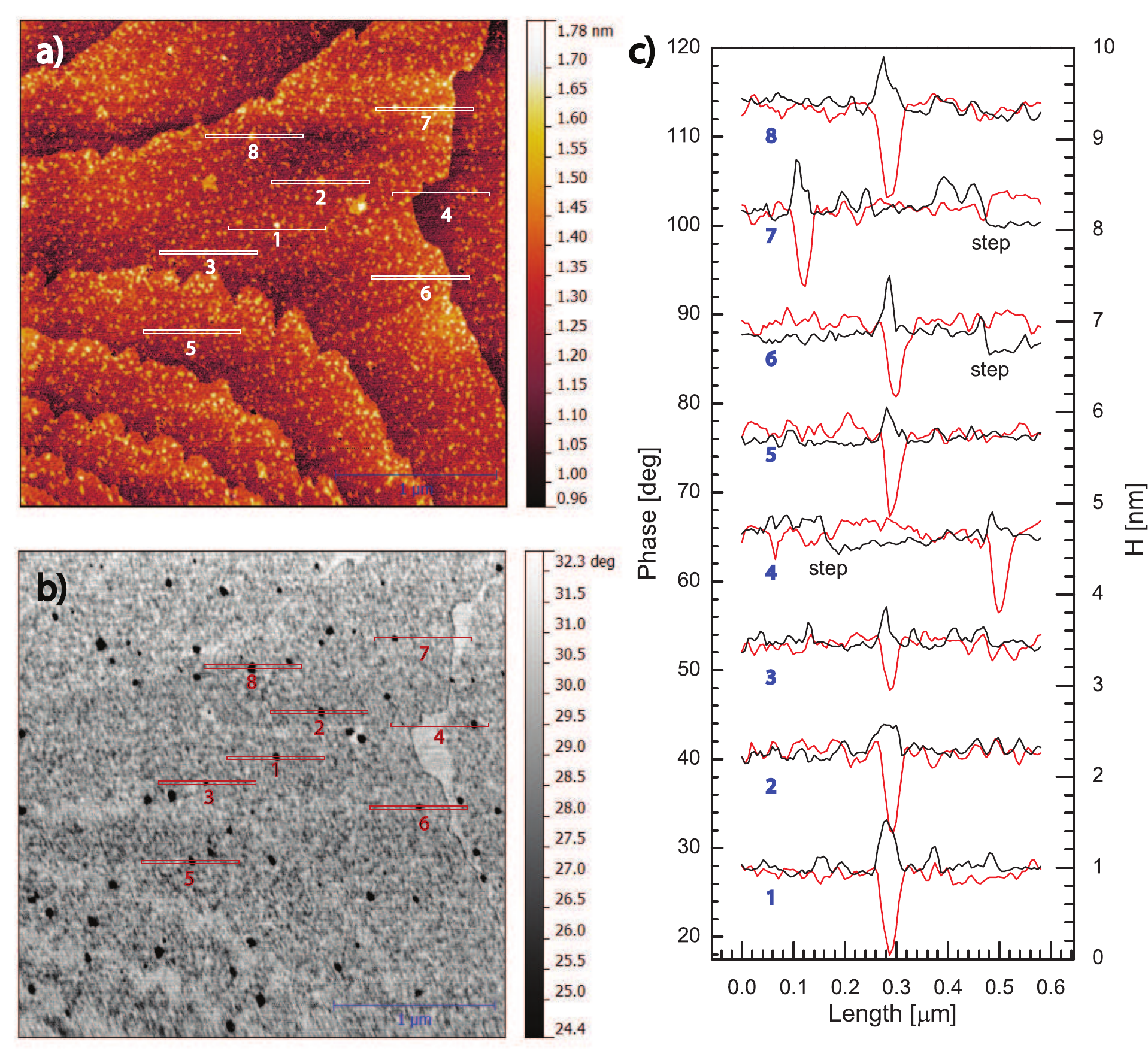}
\caption{AFM images of 3~ML Ag deposited on Pt(111) at 800~K and subsequently cooled down to room temperature. The measurements were done at ambient conditions. (a)~topography image , (b)~phase contrast image, (c)~typical line profiles trough clusters obtained from phase contrast (red line) and topography (black line) image; the profiles have been shifted vertically for sake of clarity .}
\label{fig:fig_08_bs}
\end{figure*}

Figure~\ref{fig:fig_08_bs} shows AFM data of the surface of the 3~ML Ag/Pt(111) system grown in UHV (see above) and measured under ambient conditions. The topography image of the surface presented in Fig.~\ref{fig:fig_08_bs}(a) shows step contrast which indicates that the surface is still flat. No pronounced geometric structures are observed. The phase contrast image presented in Fig.~\ref{fig:fig_08_bs}(b), reveals the presence of many small, $\sim$50~nm wide regions for which the measured phase shift is 10$^{\circ}$ lower than for the rest of the surface. We attribute these regions to the Pt clusters observed with LEEM as the number density of the clusters obtained from the AFM- and the LEEM-images is very similar (8-15 clusters/$\rm{\mu}$m$^{2}$). From the typical height profiles shown in Fig.~\ref{fig:fig_08_bs}(c) measured along the lines drawn in Fig.~\ref{fig:fig_08_bs}(a)~and~(b) we concluded that Pt clusters are about one to two layers higher than the surrounding layer. The phase shift contrast in the recorded images can be attributed to many factors~\cite{Garcia2007}, nevertheless it points to that the chemical composition of the clusters differs from the majority at the surface, which at 3~ML is mainly composed of Ag atoms. We stress that the correlation between the phase contrast and the height contrast is perfect. The black protrusions have a different chemical composition and under the applied ambient conditions probably oxidized Pt is the natural candidate. Therefore, the conclusion that the black dots consist of mainly platinum atoms is straightforward. The fact that the clusters protrude from the surface also explains the difficulty of selecting a proper focus condition in LEEM for these objects. We do believe that the footing of these clusters is still at the substrate and an oxygen layer contributes to the height differences. 

A careful look at Fig.~\ref{fig:fig_08_bs}(b) shows a pretty smooth edge on the upper terrace side of the ascending step edge. This is probably related to the initial step propagation growth mode~\cite{Jankowski2014b, Vroonhoven2005}. The area has apparently a slightly different composition and or structure. We ascribe this feature to different possibilities of dealing with stresses near a descending step.

\section{Conclusions}

The spatio-temporal \textit{in situ} information from LEEM on the growth of ultrathin Ag on Pt(111) at 750-800~K shows vivid and rich dynamical behaviour. We confirm alloying and subsequent de-alloying during the growth of the first monolayer at these elevated temperatures. Initially alloying occurs and nucleation processes are prolonged as a result of increasingly reduced mobility of Ag and Pt ad-species on the alloying surface. New nucleation events are even observed at a coverage exceeding 70\% of a monolayer! The ad-islands have a wide size distribution and are quite heterogeneous, especially the larger ones. Beyond a coverage of 50\% of a monolayer a violent segregation of Pt towards the centre of the islands occurs and bright Ag-rims become apparent. De-alloying is fast during coalescence of the adatom islands. The remaining irregularly shaped vacancy islands are quickly filled by Pt(-rich) patches (black dots) and take a compact circular shape. These features are energetically favoured since the integral step length is decreased and the boundaries between Pt(-rich) and Ag(-rich) areas are minimized. A concomitant transient relaxation of the exposed layer is deferred from the diffraction profiles. Upon further completion of the monolayer alloying re-enters and the initial pseudo-morphological structure is resumed as accompanied by re-entrant (partial) alloying, which continues in the second layer.
 
The black dots are attributed to stressed, possibly amorphous Pt(-rich) features. \textit{Ex situ} AFM experiments support this picture as the observed phase contrast images suggest chemical contrast. These Pt-features persist even after deposition of several monolayers and probably have a firm base at the Pt-substrate. 

\section{Acknowledgments}

We want to thank Robin Berkelaar for acquiring the AFM data. This work is part of ECHO research program 700.58.026, which is financed by the Chemical Sciences Division of the Netherlands Organisation for Scientific Research(NWO).

\clearpage
\bibliographystyle{iopart-num}
{\footnotesize
\bibliography{black_spots} %
}

\end{document}